\documentclass[manuscript]{acmart}

\AtBeginDocument{%
  }

\copyrightyear{2023}
\acmYear{2023}
\setcopyright{rightsretained}
\acmConference[CSCW '23 Companion]{Computer Supported Cooperative Work and Social Computing}{October 14--18, 2023}{Minneapolis, MN, USA}
\acmBooktitle{Computer Supported Cooperative Work and Social Computing (CSCW '23 Companion), October 14--18, 2023, Minneapolis, MN, USA}
\acmDOI{10.1145/3584931.3606955}
\acmISBN{979-8-4007-0129-0/23/10}

\begin{document}

\title[Large Language Models in Social Computing Research]{Shaping the Emerging Norms of Using Large Language Models in Social Computing Research}


\author{Hong Shen}
\email{hongs@cs.cmu.edu}
\affiliation{%
    \institution{Carnegie Mellon University}
    \city{Pittsburgh}
    \state{PA}
    \country{United States}
}

\author{Tianshi Li}
\email{tianshil@cs.cmu.edu}
\affiliation{%
    \institution{Carnegie Mellon University}
    \city{Pittsburgh}
    \state{PA}
    \country{United States}
}

\author{Toby Jia-Jun Li}
\email{toby.j.li@nd.edu}
\affiliation{%
    \institution{University of Notre Dame}
    \city{Notre Dame}
    \state{IN}
    \country{United States}
}

\author{Joon Sung Park}
\email{joonspk@stanford.edu}
\affiliation{%
    \institution{Stanford University}
    \city{Stanford}
    \state{CA}
    \country{United States}
}

\author{Diyi Yang}
\email{diyiy@cs.stanford.edu}
\affiliation{%
    \institution{Stanford University}
    \city{Stanford}
    \state{CA}
    \country{United States}
}

\renewcommand{\shortauthors}{Shen et al.}

\begin{abstract}
The emergence of Large Language Models (LLMs) has brought both excitement and concerns to social computing research. On the one hand, LLMs offer unprecedented capabilities in analyzing vast amounts of textual data and generating human-like responses, enabling researchers to delve into complex social phenomena. On the other hand, concerns are emerging regarding the validity, privacy, and ethics of the research when LLMs are involved. This SIG aims at offering an open space for social computing researchers who are interested in understanding the impacts of LLMs to discuss their current practices, perspectives, challenges when engaging with LLMs in their everyday work and collectively shaping the emerging norms of using LLMs in social computing research.
\end{abstract}



\maketitle
\section{Background}
The development of large language models (LLMs) such as ChatGPT has brought both excitements and concerns to the field of social computing. On the one hand, LLMs present important opportunities that leverage the vast amount of human behavioral data captured in the model~\cite{park2023generative} to analyze and augment interactions in social computing systems. For instance, LLM-powered tools have been shown to assist researchers to more efficiently analyze textual data~\cite{xiao2023supporting, ziems2023can}, replicate social science studies~\cite{Binz_2023}, and enable new ways to prototype emergent social dynamics in social computing systems where it is infeasible or dangerous to conduct in the wild studies~\cite{park2022socialsimulacra}. 

On the other hand, concerns about applying LLMs to social computing research~\cite{bail2023can} that parallel previous discussions on privacy and consent in computational social science research \cite{lewis2008tastes, salganik2019bit} have also arisen. In particular, we focus on three key themes: validity, privacy and ethics. Can we ensure the validity of findings generated by a non-deterministic black-box model whose output may depend on the nuances captured in the input prompts? Can we protect the privacy of the subjects whose data maybe captured in the models' training data? And can we put in place proper guardrails that will encourage ethical application of LLMs in social computing research in the face of their risks, including but not limited to their potential misuse of LLMs for manipulation or deception? 

The primary goal of this Special Interest Group (SIG) is to provide an inclusive platform for social computing researchers who wish to explore the implications of LLMs in their day-to-day research activities. In particular, we aim to explore the following questions: 
\begin{itemize}
\item How to better design online studies to effectively prevent LLM-based spams? For example, how to prevent, recognize and filter out spammers who use LLMs to fill out online surveys?
\item How to utilize LLMs to analyze human-generated data and effectively evaluate their performance? For example, how to construct ground truth when using LLMs in analyzing qualitative data (e.g., open-ended survey responses)? What are the evaluation metrics should we use (e.g., shall we calculate inter-coder reliability)? 
\item How to accurately document the utilization of LLMs in research while simultaneously acknowledging its inherent limitations and biases in an effective manner? 
\item How to preserve the privacy of the study data when we use LLMs in data analysis? For example, how to effectively remove Personal Identifiable Information (PII) from survey responses, interview transcripts, and data scraped from the web?
\item How to address ethical concerns related with using LLMs in social computing? For example, how to craft informed consent to inform study participants the potential of using LLMs in the study design? How to ethically use dataset containing real-world LLMs usage data in research?
\item How to mitigate the equity concerns associated with the substantial cost, computational resources, and technical expertise required for employing LLMs, considering the unequal access to these resources among different research teams? 
\end{itemize}

In an attempt to ground the above questions in a more concrete manner, below we discuss the impacts of LLMs in social computing research on the following four aspects: Data collection, data generation, data analysis as well as system deployment and evaluation. 

\section{The impact of LLMs on data collection}
Social computing researchers are already experiencing the impacts of LLMs in their work. One particular area is during the data collection stage. On the one hand, the rapidly advancing capabilities of LLMs to mimic human behaviors have presented numerous promising opportunities. For instance, researchers can leverage LLMs to generate hypothetical scenarios (e.g., vignettes) to collect data from human participants. Additionally, LLM-based agents can also be introduced into multiplayer games, opening up new avenues for studying human behavior in interactive settings \cite{bail2023can}.
However, these capabilities also give rise to a variety of concerns. For example, researchers have shown that chatbots powered by LLMs can effectively mimic survey respondents with diverse backgrounds \cite{argyle2023ai}. This poses a challenge for researchers engaged in online studies, as it becomes increasingly difficult to differentiate between AI-based spammers/bots from genuine human participants. Traditional methods used to identify and filter out spammers may no longer be effective in this context. Moreover, the introduction of LLMs in different parts of the research design, including using LLMs to analyze human-generated data and/or using LLMs to simulate human behavior, will also likely need an update on the consent process. What is the best practices to craft informed consent with human participants when LLMs is involved? 

\section{The use of LLMs in generating synthetic data}
LLMs capture the human behaviors that are represented in their training data~\cite{Binz_2023, park2022socialsimulacra}, and as such, these models can replicate these behaviors when prompted. Recent studies have shown that the human behavior generated by these models is qualitatively believable~\cite{park2023generative} and, at times, accurate enough to replicate some social science studies~\cite{ziems2023can} and surveys~\cite{argyle2023ai}. This capacity for the model to generate human behavior offers opportunities to enable new ways of studying and augmenting social computing systems. For instance, these models can allow the designers of a social system to prototype the social dynamics that only emerge at scale, to iterate without exposing the users to potentially flawed system design. They can also bring about new ways of conducting computational social science by replicating results that were only achievable via crowd participants or empirical studies. However, the applications of LLMs inherit the imperfections and biases of the underlying model. Their output might depend on the subtle nuances of a prompt, while their biases might misrepresent certain populations. We posit that our community will need to continuously validate and benchmark the use of LLMs in social computing while emphasizing the importance of directly connecting with human stakeholders.

\section{The use of LLMs in analyzing data}
At the data analysis stage of social computing research, we are witnessing emerging use of LLMs (and recently, \textit{large} pre-trained ones) in both qualitative and quantitative methods. 

In qualitative research, academic tools such as PaTAT~\cite{gebreegziabher2023patat} and CollabCoder~\cite{gao2023collabcoder} as well as commercial products such as the AI Coding feature in ATLAS.ti\footnote{\url{https://atlasti.com/ai-coding}} utilize LLMs to aid in the qualitative coding process for textual data. Generally, these tools employ LLMs to analyze data, propose new codes for inductive coding procedures, understand the semantics of codes as users create and assign them, and suggest codes for data items. They also assist with the sensemaking process by summarizing, synthesizing, aggregating, or visualizing data.

For quantitative methods, AI and LLM-enabled assistants and ``pair programmers'' have been developed to recommend analyses and statistical procedures based on data characteristics, conduct data queries, and implement data analysis code in response to user prompts in exploratory data science scenarios~\cite{mcnutt2023design, wang2019human, ning2023empirical}. Commercial products such as Tableau AI\footnote{\url{https://www.tableau.com/solutions/ai-analytics}} suggest metrics based on the data domain, generate insights, and allow users to ask questions about the underlying data using natural language.

Despite their potential in improving the efficiency of the analysis process, facilitating additional insight discovery, and reducing learning curves, the use of LLMs also presents new challenges and raises concerns regarding their application in social computing research. For instance, LLMs have been found to display biases and stereotypes in their outputs~\cite{liang2021towards,nadeem-etal-2021-stereoset}, which could influence the data analysis process, especially in domains where understanding the socio-cultural context is essential for data interpretation. Moreover, the collaboration between human researchers and LLM-enabled tools in analysis tasks poses challenges in ensuring user autonomy, preventing over-reliance, and promoting effective human learning about data and patterns, this challenge can be amplified by the lack of interpretability in LLMs~\cite{gebreegziabher2023patat}. The application of LLMs in data analysis introduces additional data privacy challenges since many LLMs lack transparency regarding their usage of user data. This raises questions about creating informed consent protocols to notify participants about the use of LLMs in analyzing their data. Furthermore, the community needs new guidelines and norms regarding evaluation metrics when LLMs are used alongside human coders in data analysis.


\section{The methods for deploying/studying LLM-enabled socio-technical systems}

The success of LLMs is going to have a profound impact on socio-technical systems in various domains that humans use natural languages to interact with.
Over the past few months, news articles have covered systems built with LLMs used for mental health support\footnote{\url{https://gizmodo.com/mental-health-therapy-app-ai-koko-chatgpt-rob-morris-1849965534}},  education\footnote{\url{https://fortune.com/2023/02/22/chatgpt-ai-openai-educatoin-tutor-teaching-school/}},  legal services\footnote{\url{https://gizmodo.com/donotpay-speeding-ticket-chatgpt-1849960272}},  job searching advice\footnote{\url{https://www.forbes.com/sites/jackkelly/2023/04/03/how-to-leverage-ai-and-use-chatgpt-in-your-job-search-according-to-rsum-writers-and-career-coaches/?sh=728117a5ac5a}}, and many other purposes.
While they demonstrate exciting opportunities of advancing these fields, concerns and backlashes have been raised by the public.
For example, in January 2023, a company called Koko that offers mental health services tested responses generated by GPT-3 on thousands of its users.
A co-founder tweeted about their experiments and it soon sparked a heated discussion around the ethics of this research.
People questioned their informed consent process, the legitimacy of testing an unproven technology on real users, and even the appropriateness of involving AI in such a process at all.
Although the co-founder later clarified that Koko users knew the messages were co-written by a bot, it did not resolve all these concerns.

Researching systems built with LLMs is a delicate process.
However, the lack of clear guidelines for conducting research in this field will affect both researchers who design, develop, and deploy socio-technical systems powered by LLMs, and researchers who conduct empirical studies to investigate how real-world users interact with these systems.
In this SIG, we aim to take the first step towards the development of the guidelines.
The questions that need in-depth discussions include but are not limited to the following.
For researchers who aim to deploy a novel LLM-enabled system, how should they determine whether AI-based intervention is appropriate for the selected use case?
How should they disclose the use of LLMs to their users?
Deploying certain services (e.g., mental health support) may inevitably lead the users to expose sensitive information about themselves and other people (i.e., interdependent privacy~\cite{humbert2019survey}).
How should researchers process traces from the study that may involve such sensitive information?
Relatedly, the natural language interfaces give users a great amount of flexibility, which means the LLM-enabled services may not have definite use scenarios (e.g., ChatGPT) or the users may use them in unexpected ways.
Hence, researchers who want to study the use of LLM-enabled systems are facing challenges in handling unexpected privacy harms (e.g., economical, reputational, psychological harms~\cite{citron2022privacy}), which may affect the choice of tools for analysis (e.g., local vs. cloud-based tools).


\section{Conclusion}

The rapid development of Large Language Models (LLMs) has already had a significant impact on various aspects of social computing research, encompassing areas such as data collection, data generation, data analysis, as well as system deployment and evaluation. However, alongside the excitement surrounding these advancements, concerns have emerged regarding issues of validity, privacy, and ethics. This Special Interest Group (SIG) aims to provide a much-needed space for researchers who are interested in comprehending the impacts of LLMs on their work. It offers an opportunity for the members of the community to openly discuss their current practices, perspectives, and challenges when engaging with LLMs in their day-to-day activities and collectively shaping the emerging norms of LLMs-impacted social computing research.



\bibliographystyle{ACM-Reference-Format}
\bibliography{LLM}


\begin{thebibliography}{18}


\ifx \showCODEN    \undefined \def \showCODEN     #1{\unskip}     \fi
\ifx \showDOI      \undefined \def \showDOI       #1{#1}\fi
\ifx \showISBNx    \undefined \def \showISBNx     #1{\unskip}     \fi
\ifx \showISBNxiii \undefined \def \showISBNxiii  #1{\unskip}     \fi
\ifx \showISSN     \undefined \def \showISSN      #1{\unskip}     \fi
\ifx \showLCCN     \undefined \def \showLCCN      #1{\unskip}     \fi
\ifx \shownote     \undefined \def \shownote      #1{#1}          \fi
\ifx \showarticletitle \undefined \def \showarticletitle #1{#1}   \fi
\ifx \showURL      \undefined \def \showURL       {\relax}        \fi
\providecommand\bibfield[2]{#2}
\providecommand\bibinfo[2]{#2}
\providecommand\natexlab[1]{#1}
\providecommand\showeprint[2][]{arXiv:#2}

\bibitem[Argyle et~al\mbox{.}(2023)]%
        {argyle2023ai}
\bibfield{author}{\bibinfo{person}{Lisa~P Argyle}, \bibinfo{person}{Ethan
  Busby}, \bibinfo{person}{Joshua Gubler}, \bibinfo{person}{Chris Bail},
  \bibinfo{person}{Thomas Howe}, \bibinfo{person}{Christopher Rytting}, {and}
  \bibinfo{person}{David Wingate}.} \bibinfo{year}{2023}\natexlab{}.
\newblock \showarticletitle{AI Chat Assistants can Improve Conversations about
  Divisive Topics}.
\newblock \bibinfo{journal}{\emph{arXiv preprint arXiv:2302.07268}}
  (\bibinfo{year}{2023}).
\newblock


\bibitem[Bail(2023)]%
        {bail2023can}
\bibfield{author}{\bibinfo{person}{Christopher~A Bail}.}
  \bibinfo{year}{2023}\natexlab{}.
\newblock \showarticletitle{Can Generative AI Improve Social Science?}
\newblock  (\bibinfo{year}{2023}).
\newblock
\urldef\tempurl%
\url{https://osf.io/preprints/socarxiv/rwtzs}
\showURL{%
\tempurl}


\bibitem[Binz and Schulz(2023)]%
        {Binz_2023}
\bibfield{author}{\bibinfo{person}{Marcel Binz} {and} \bibinfo{person}{Eric
  Schulz}.} \bibinfo{year}{2023}\natexlab{}.
\newblock \showarticletitle{Using cognitive psychology to understand {GPT}-3}.
\newblock \bibinfo{journal}{\emph{Proceedings of the National Academy of
  Sciences}} \bibinfo{volume}{120}, \bibinfo{number}{6} (\bibinfo{date}{feb}
  \bibinfo{year}{2023}).
\newblock
\urldef\tempurl%
\url{https://doi.org/10.1073/pnas.2218523120}
\showDOI{\tempurl}


\bibitem[Citron and Solove(2022)]%
        {citron2022privacy}
\bibfield{author}{\bibinfo{person}{Danielle~Keats Citron} {and}
  \bibinfo{person}{Daniel~J Solove}.} \bibinfo{year}{2022}\natexlab{}.
\newblock \showarticletitle{Privacy harms}.
\newblock \bibinfo{journal}{\emph{BUL Rev.}}  \bibinfo{volume}{102}
  (\bibinfo{year}{2022}), \bibinfo{pages}{793}.
\newblock


\bibitem[Gao et~al\mbox{.}(2023)]%
        {gao2023collabcoder}
\bibfield{author}{\bibinfo{person}{Jie Gao}, \bibinfo{person}{Yuchen Guo},
  \bibinfo{person}{Gionnieve Lim}, \bibinfo{person}{Tianqin Zhan},
  \bibinfo{person}{Zheng Zhang}, \bibinfo{person}{Toby Jia-Jun Li}, {and}
  \bibinfo{person}{Simon~Tangi Perrault}.} \bibinfo{year}{2023}\natexlab{}.
\newblock \showarticletitle{CollabCoder: A GPT-Powered Workflow for
  Collaborative Qualitative Analysis}.
\newblock \bibinfo{journal}{\emph{arXiv preprint arXiv:2304.07366}}
  (\bibinfo{year}{2023}).
\newblock


\bibitem[Gebreegziabher et~al\mbox{.}(2023)]%
        {gebreegziabher2023patat}
\bibfield{author}{\bibinfo{person}{Simret~Araya Gebreegziabher},
  \bibinfo{person}{Zheng Zhang}, \bibinfo{person}{Xiaohang Tang},
  \bibinfo{person}{Yihao Meng}, \bibinfo{person}{Elena~L. Glassman}, {and}
  \bibinfo{person}{Toby Jia-Jun Li}.} \bibinfo{year}{2023}\natexlab{}.
\newblock \showarticletitle{PaTAT: Human-AI Collaborative Qualitative Coding
  with Explainable Interactive Rule Synthesis}. In
  \bibinfo{booktitle}{\emph{Proceedings of the 2023 CHI Conference on Human
  Factors in Computing Systems}} (Hamburg, Germany) \emph{(\bibinfo{series}{CHI
  '23})}. \bibinfo{publisher}{Association for Computing Machinery},
  \bibinfo{address}{New York, NY, USA}, Article \bibinfo{articleno}{362},
  \bibinfo{numpages}{19}~pages.
\newblock
\showISBNx{9781450394215}
\urldef\tempurl%
\url{https://doi.org/10.1145/3544548.3581352}
\showDOI{\tempurl}


\bibitem[Humbert et~al\mbox{.}(2019)]%
        {humbert2019survey}
\bibfield{author}{\bibinfo{person}{Mathias Humbert}, \bibinfo{person}{Benjamin
  Trubert}, {and} \bibinfo{person}{K{\'e}vin Huguenin}.}
  \bibinfo{year}{2019}\natexlab{}.
\newblock \showarticletitle{A survey on interdependent privacy}.
\newblock \bibinfo{journal}{\emph{ACM Computing Surveys (CSUR)}}
  \bibinfo{volume}{52}, \bibinfo{number}{6} (\bibinfo{year}{2019}),
  \bibinfo{pages}{1--40}.
\newblock


\bibitem[Lewis et~al\mbox{.}(2008)]%
        {lewis2008tastes}
\bibfield{author}{\bibinfo{person}{Kevin Lewis}, \bibinfo{person}{Jason
  Kaufman}, \bibinfo{person}{Marco Gonzalez}, \bibinfo{person}{Andreas Wimmer},
  {and} \bibinfo{person}{Nicholas Christakis}.}
  \bibinfo{year}{2008}\natexlab{}.
\newblock \showarticletitle{Tastes, ties, and time: A new social network
  dataset using Facebook. com}.
\newblock \bibinfo{journal}{\emph{Social networks}} \bibinfo{volume}{30},
  \bibinfo{number}{4} (\bibinfo{year}{2008}), \bibinfo{pages}{330--342}.
\newblock


\bibitem[Liang et~al\mbox{.}(2021)]%
        {liang2021towards}
\bibfield{author}{\bibinfo{person}{Paul~Pu Liang}, \bibinfo{person}{Chiyu Wu},
  \bibinfo{person}{Louis-Philippe Morency}, {and} \bibinfo{person}{Ruslan
  Salakhutdinov}.} \bibinfo{year}{2021}\natexlab{}.
\newblock \showarticletitle{Towards understanding and mitigating social biases
  in language models}. In \bibinfo{booktitle}{\emph{International Conference on
  Machine Learning}}. PMLR, \bibinfo{pages}{6565--6576}.
\newblock


\bibitem[Mcnutt et~al\mbox{.}(2023)]%
        {mcnutt2023design}
\bibfield{author}{\bibinfo{person}{Andrew~M Mcnutt}, \bibinfo{person}{Chenglong
  Wang}, \bibinfo{person}{Robert~A Deline}, {and} \bibinfo{person}{Steven~M.
  Drucker}.} \bibinfo{year}{2023}\natexlab{}.
\newblock \showarticletitle{On the Design of AI-Powered Code Assistants for
  Notebooks}. In \bibinfo{booktitle}{\emph{Proceedings of the 2023 CHI
  Conference on Human Factors in Computing Systems}} (Hamburg, Germany)
  \emph{(\bibinfo{series}{CHI '23})}. \bibinfo{publisher}{Association for
  Computing Machinery}, \bibinfo{address}{New York, NY, USA}, Article
  \bibinfo{articleno}{434}, \bibinfo{numpages}{16}~pages.
\newblock
\showISBNx{9781450394215}
\urldef\tempurl%
\url{https://doi.org/10.1145/3544548.3580940}
\showDOI{\tempurl}


\bibitem[Nadeem et~al\mbox{.}(2021)]%
        {nadeem-etal-2021-stereoset}
\bibfield{author}{\bibinfo{person}{Moin Nadeem}, \bibinfo{person}{Anna Bethke},
  {and} \bibinfo{person}{Siva Reddy}.} \bibinfo{year}{2021}\natexlab{}.
\newblock \showarticletitle{{S}tereo{S}et: Measuring stereotypical bias in
  pretrained language models}. In \bibinfo{booktitle}{\emph{Proceedings of the
  59th Annual Meeting of the Association for Computational Linguistics and the
  11th International Joint Conference on Natural Language Processing (Volume 1:
  Long Papers)}}. \bibinfo{publisher}{Association for Computational
  Linguistics}, \bibinfo{address}{Online}, \bibinfo{pages}{5356--5371}.
\newblock
\urldef\tempurl%
\url{https://doi.org/10.18653/v1/2021.acl-long.416}
\showDOI{\tempurl}


\bibitem[Ning et~al\mbox{.}(2023)]%
        {ning2023empirical}
\bibfield{author}{\bibinfo{person}{Zheng Ning}, \bibinfo{person}{Zheng Zhang},
  \bibinfo{person}{Tianyi Sun}, \bibinfo{person}{Yuan Tian},
  \bibinfo{person}{Tianyi Zhang}, {and} \bibinfo{person}{Toby Jia-Jun Li}.}
  \bibinfo{year}{2023}\natexlab{}.
\newblock \showarticletitle{An Empirical Study of Model Errors and User Error
  Discovery and Repair Strategies in Natural Language Database Queries}. In
  \bibinfo{booktitle}{\emph{Proceedings of the 28th International Conference on
  Intelligent User Interfaces}} (Sydney, NSW, Australia)
  \emph{(\bibinfo{series}{IUI '23})}. \bibinfo{publisher}{Association for
  Computing Machinery}, \bibinfo{address}{New York, NY, USA},
  \bibinfo{pages}{633–649}.
\newblock
\showISBNx{9798400701061}
\urldef\tempurl%
\url{https://doi.org/10.1145/3581641.3584067}
\showDOI{\tempurl}


\bibitem[Park et~al\mbox{.}(2023)]%
        {park2023generative}
\bibfield{author}{\bibinfo{person}{Joon~Sung Park}, \bibinfo{person}{Joseph~C.
  O'Brien}, \bibinfo{person}{Carrie~J. Cai}, \bibinfo{person}{Meredith~Ringel
  Morris}, \bibinfo{person}{Percy Liang}, {and} \bibinfo{person}{Michael~S.
  Bernstein}.} \bibinfo{year}{2023}\natexlab{}.
\newblock \bibinfo{title}{Generative Agents: Interactive Simulacra of Human
  Behavior}.
\newblock
\newblock
\showeprint[arxiv]{2304.03442}~[cs.HC]


\bibitem[Park et~al\mbox{.}(2022)]%
        {park2022socialsimulacra}
\bibfield{author}{\bibinfo{person}{Joon~Sung Park}, \bibinfo{person}{Lindsay
  Popowski}, \bibinfo{person}{Carrie~J. Cai}, \bibinfo{person}{Meredith~Ringel
  Morris}, \bibinfo{person}{Percy Liang}, {and} \bibinfo{person}{Michael~S.
  Bernstein}.} \bibinfo{year}{2022}\natexlab{}.
\newblock \showarticletitle{Social Simulacra: Creating Populated Prototypes for
  Social Computing Systems}. In \bibinfo{booktitle}{\emph{In the 35th Annual
  ACM Symposium on User Interface Software and Technology (UIST '22)}} (Bend,
  OR, USA) \emph{(\bibinfo{series}{UIST '22})}. \bibinfo{publisher}{Association
  for Computing Machinery}, \bibinfo{address}{New York, NY, USA}.
\newblock
\showISBNx{9781450393201}
\urldef\tempurl%
\url{https://doi.org/10.1145/3526113.3545616}
\showDOI{\tempurl}


\bibitem[Salganik(2019)]%
        {salganik2019bit}
\bibfield{author}{\bibinfo{person}{Matthew~J Salganik}.}
  \bibinfo{year}{2019}\natexlab{}.
\newblock \bibinfo{booktitle}{\emph{Bit by bit: Social research in the digital
  age}}.
\newblock \bibinfo{publisher}{Princeton University Press}.
\newblock


\bibitem[Wang et~al\mbox{.}(2019)]%
        {wang2019human}
\bibfield{author}{\bibinfo{person}{Dakuo Wang}, \bibinfo{person}{Justin~D.
  Weisz}, \bibinfo{person}{Michael Muller}, \bibinfo{person}{Parikshit Ram},
  \bibinfo{person}{Werner Geyer}, \bibinfo{person}{Casey Dugan},
  \bibinfo{person}{Yla Tausczik}, \bibinfo{person}{Horst Samulowitz}, {and}
  \bibinfo{person}{Alexander Gray}.} \bibinfo{year}{2019}\natexlab{}.
\newblock \showarticletitle{Human-AI Collaboration in Data Science: Exploring
  Data Scientists' Perceptions of Automated AI}.
\newblock \bibinfo{journal}{\emph{Proc. ACM Hum.-Comput. Interact.}}
  \bibinfo{volume}{3}, \bibinfo{number}{CSCW}, Article \bibinfo{articleno}{211}
  (\bibinfo{date}{nov} \bibinfo{year}{2019}), \bibinfo{numpages}{24}~pages.
\newblock
\urldef\tempurl%
\url{https://doi.org/10.1145/3359313}
\showDOI{\tempurl}


\bibitem[Xiao et~al\mbox{.}(2023)]%
        {xiao2023supporting}
\bibfield{author}{\bibinfo{person}{Ziang Xiao}, \bibinfo{person}{Xingdi Yuan},
  \bibinfo{person}{Q~Vera Liao}, \bibinfo{person}{Rania Abdelghani}, {and}
  \bibinfo{person}{Pierre-Yves Oudeyer}.} \bibinfo{year}{2023}\natexlab{}.
\newblock \showarticletitle{Supporting Qualitative Analysis with Large Language
  Models: Combining Codebook with GPT-3 for Deductive Coding}. In
  \bibinfo{booktitle}{\emph{Companion Proceedings of the 28th International
  Conference on Intelligent User Interfaces}}. \bibinfo{pages}{75--78}.
\newblock


\bibitem[Ziems et~al\mbox{.}(2023)]%
        {ziems2023can}
\bibfield{author}{\bibinfo{person}{Caleb Ziems}, \bibinfo{person}{William
  Held}, \bibinfo{person}{Omar Shaikh}, \bibinfo{person}{Jiaao Chen},
  \bibinfo{person}{Zhehao Zhang}, {and} \bibinfo{person}{Diyi Yang}.}
  \bibinfo{year}{2023}\natexlab{}.
\newblock \showarticletitle{Can Large Language Models Transform Computational
  Social Science?}
\newblock \bibinfo{journal}{\emph{arXiv preprint arXiv:2305.03514}}
  (\bibinfo{year}{2023}).
\newblock


\end{thebibliography}


\end{document}